\documentstyle[12pt]{article}
\def\be{\begin{equation}}
\def\ee{\end{equation}}

\begin{document}
\begin{titlepage}

\title{The old frequency decomposition problem in the light
       of new quantization methods 
    \thanks{Talk given at the Eighth Marcel Grossmann meeting,
                    Jerusalem, June 22--27, 1997.
                     }
                          }

\author{Franz Embacher\\
        Institut f\"ur Theoretische Physik\\
        Universit\"at Wien\\
        Boltzmanngasse 5\\
        A-1090 Wien\\
        \\
        E-mail: fe@pap.univie.ac.at\\
        \\
        UWThPh-1997-21\\
        gr-qc/9708015
                      } 
\date{}

\maketitle

\begin{abstract}
The question is raised whether the unique 
decomposition of the physical Hilbert space, as emerging in 
the refined algebraic quantization of a constrained system,  
may be understood in terms of the old Klein-Gordon type 
quantization. 
\end{abstract}

\end{titlepage}

\section{Two quantization methods and the issue of 
 frequenc\-y decomposition}

The conventional (``old'') method to quantize a 
theory containing a momentum squared constraint is inspired from
the situation of a (scalar) particle moving in a space-time 
background. 
Restricting ourselves to finite dimensional systems, the wave 
equation (minisuperspace Wheeler-DeWitt equation) is of the type 
\be
{\cal C}\,\psi \equiv 
(-\nabla_\alpha \nabla^\alpha + U )\,\psi = 0\, , 
\label{WDW}
\ee
with $\nabla_\alpha \nabla^\alpha$ being the Laplacian with respect to
some Pseudo-Riemannian metric $g_{\alpha\beta}$ on a finite 
dimensional manifold ${\cal M}$, the real function $U$ playing 
the role of a potential. 
The space of sufficiently well-behaved solutions 
admits the well-known indefinite Klein-Gordon type scalar product 
\be
Q(\psi_1,\psi_2)= -\,\frac{i}{2}\, \int_\Sigma d\Sigma^\alpha\,
(\psi_1^*\stackrel{\leftrightarrow}{\nabla}_\alpha \psi_2) \, , 
\label{KG}
\ee
where $\Sigma$ is a spacelike hypersurface 
(with sufficiently regular asymptotic behaviour). 
If the background structure 
$({\cal M},g_{\alpha\beta},U)$ admits a local symmetry with timelike 
trajectories (as e.g. for flat $g_{\alpha\beta}$ and 
constant non-negative $U$, 
in which case (\ref{WDW}) is the Klein-Gordon equation)
there is a unique decomposition of wave functions 
into positive and negative frequency modes 
($Q$ restricted to the positive/negative frequency sector 
being a positive/negative definite scalar product, and the two 
sectors being orthogonal to each other). 
In the case of a generic background, it is common folklore that there 
exists no such unique decomposition 
\cite{Kuchar}. 
\medskip 

However, there is another quantization scheme for the same sort 
of systems that starts from the inner product 
\begin{equation}
\langle\psi_1,\psi_2\rangle 
=\int_{\cal M} d^n x\,\sqrt{-g}\,\psi_1^*\psi_2 
\label{KIN}
\end{equation} 
on the Hilbert space of square integrable functions on the manifold. 
The basic idea of how to proceed dates back to the Sixties 
(the earliest reference I am aware of is Nachtmann 
\cite{Nachtmann}), 
but there does not seem to have emerged a tradition from that 
(see however Refs. \cite{RumpfUrbantke}). 
After a re-invention of the ansatz 
in the Nineties 
\cite{Higuchi,Landsmanetal,Marolf}, 
this approach has been developed further and has become a viable 
method by which the quantization of full general relativity is currently 
being attacked \cite{Aetal}. 
It runs under several names, the best known being 
``refined algebraic quantization'' (others being 
''Rieffel induction'' and ''group averaging''). 
\medskip 

Without going into the details, I just summarize that (\ref{KIN}) 
gives rise to a positive definite inner product 
$\langle\, , \,\rangle_{{}_{\rm phys}}$ on a suitably defined 
set of solutions of (\ref{WDW}). 
(When inserting two solutions of (\ref{WDW}) 
into (\ref{KIN}), one would in general 
obtain an infinite result, but if the wave operator $\cal C$ in 
(\ref{WDW}) is self-adjoint, this can be cured by ``dividing by an 
infinite constant'' or, more precise, by averaging over the group 
generated by $\cal C$). Thus, upon completion, 
one ends up with what is usually called 
the physical Hilbert space 
$({\cal H}_{{}_{\rm phys}}, \langle\, , \,\rangle_{{}_{\rm phys}})$. 
\medskip 

Most researchers employing this new scheme simply forget about the 
structure (\ref{KG}) that has played an important role in the early 
years of quantum gravity. Since in quantum gravity or quantum cosmology, 
(mini)superspace plays a role fundamentally different from the 
space-time manifold in the particle quantization problem, 
one may take the point of view that the notion of positive and
negative frequency modes does not play any substantial role there. 
However, the structure (\ref{KG}) still exists, even if it is 
not payed any attention. 
(I ignore here the problem that (\ref{KG}) is ill-defined in the 
full superspace context and must be regularized. 
In the framework under consideration, $Q$ is well-defined 
on ${\cal H}_{{}_{\rm phys}}$.) 
\medskip 

What can we infer from the fact that $Q$ and 
$\langle\, , \,\rangle_{{}_{\rm phys}}$ 
{\it coexist} on one and the same 
space? The former quantity may be represented 
in terms of the latter by 
$Q(\psi_1,\psi_2) = \langle \psi_1, K \psi_2\rangle_{{}_{\rm phys}}$, 
where $K$ is a linear (supposedly self-adjoint) operator. 
In reasonable cases, its 
(generalized) eigenvalues come in pairs
$(-\lambda,\lambda)$ off zero (in the case of the Klein-Gordon equation
we have even $K^2=1$), so that the Hilbert space uniquely decomposes as 
${\cal H}_{{}_{\rm phys}} = {\cal H}^+ \oplus {\cal H}^-$. 
Moreover, $Q$ is positive/negative definite on ${\cal H}^\pm$ and
the two subspaces are orthogonal to each other with respect to both scalar 
products. This decomposition has been singled out by the 
global structure of 
$({\cal M},g_{\alpha\beta},U)$ (note that no
local symmetry is necessary for the construction to work). 
Recently, Hartle and Marolf have exploited the coexistence
of the two scalar products, though with different motivation 
\cite{HartleMarolf}. 

\section{Understanding new issues in terms of old 
$\,\,\,\,\,\,\,\,\,$ methods?}

Can the decomposition defined above be viewed as ``the correct'' 
identification of positive and negative 
frequencies? Note that the refined algebraic quantization scheme 
provides a structure that 
is in a sense invisible for the Klein-Gordon type approach 
{\it although it does not need any additional input}. 
Perhaps a clarification of this situation could improve our 
understanding of what happens when we quantize a constrained system. 
I cannot resolve this puzzle, but I would like to 
mention a candidate for a procedure defined 
within the Klein-Gordon framework 
but transcending the differential geometric 
setting. It might possibly show us a way how to make contact between 
the two methods. 
\medskip 

In Ref. \cite{FE}, a framework for treating quite general wave equations
of the type (\ref{WDW}) with positive potential was proposed. 
Writing the wave function as $\psi = \chi D e^{i S}$, 
(with $S$ being a sufficiently globally regular 
solution of the classical Hamilton-Jacobi 
equation and $D$ a real function satisfying a certain 
conservation equation), the wave equation (\ref{WDW}) reads
$i\, \partial_t \chi = (\,\frac{1}{2}\,\partial_{tt} + h)\chi$. 
Here $t=-S$ and $h$ is a linear differential operator acting 
tangential to the hypersurfaces $\Sigma_t$ of constant $t$. 
Although resembling a WKB scheme, no approximation 
is applied so far. 
In Ref. \cite{FE} it is argued that if an operator 
$H$ is defined as a series of the type 
\be
H = h -\,\frac{1}{2}\, h^2-\,\frac{i}{2}\,\dot{h} + 
 \frac{1}{2}\, h^3 +\frac{i}{2}\,\{h,\dot{h}\}-\, 
\frac{1}{4}\, \ddot{h} + \dots 
\label{iter}
\ee
(as emerging from the iteration of a certain differential equation), 
where $\dot{h}\equiv [ \partial_t, h]$, then 
any solution of the Schr{\"o}dinger equation 
\be 
i\,\partial_t \chi = H \chi 
\label{SCHR}
\ee 
solves (\ref{WDW}). 
The actual convergence of (\ref{iter}) seems to depend 
on the particular background $({\cal M},g_{\alpha\beta},U)$, and
it is here that some models might be excluded (while 
remaining intact from the point of view of differential geometry). 
In case of convergence (which has been checked for the simple 
cases $h=a + b t + c t^2$, $h=\alpha/t$ and 
$h=\beta/t^2$ (this last one having applications to 
FRW quantum cosmology), 
the set of solutions $\psi$ obtained in this way 
forms a subspace ${\cal H}^{\prime +}$ which is independent of the 
choice of the pair $(S,D)$ --- called a ``WKB-branch'' --- 
in which it is calculated and, together with 
its complex conjugate ${\cal H}^{\prime -}$, decomposes 
${\cal H}_{{}_{\rm phys}}$ into a direct orthogonal sum.
$Q$ is positive/negative definite on ${\cal H}^{\prime\pm}$ 
just as was the case for ${\cal H}^\pm$ above.  
\medskip 

I cannot answer the natural question arising here, whether 
${\cal H}^{\prime\pm}$ 
have something to do (or are even identical) with 
${\cal H}^\pm$ (except for the flat space Klein-Gordon 
equation, where they {\it are} identical). 
Maybe pursuing this route could clarify 
why the decomposition based on ${\cal H}^\pm$ is invisible to 
the Klein-Gordon quantization scheme, at least as long as one remains 
within the pure differential geometric framework.

\end{document}